\documentclass[cameraready]{Interspeech}
\title{L-Proto: Language-Aware Episodic Prototypical Training for Multilingual Speaker Verification}

\author[affiliation={}, orcid=0000-0001-7229-8123]{Hyung-Seok}{Oh}
\author[affiliation={}, orcid=0009-0002-4673-9882]{Deok-Hyeon}{Cho}
\author[affiliation={}, orcid=0000-0002-2287-9111]{Seung-Bin}{Kim}
\author[affiliation={}, orcid=0000-0002-6249-4996, correspondingauthor]{Seong-Whan}{Lee}
\address{
    Department of Artificial Intelligence, Korea University, Seoul, Korea
}

\email{hs\_oh@korea.ac.kr, sb-kim@korea.ac.kr, dh\_cho@korea.ac.kr, sw.lee@korea.ac.kr}


\usepackage{comment}
\usepackage{amsmath}
\usepackage{graphicx}
\usepackage{subcaption}
\usepackage{algorithm}
\usepackage{algorithmic}
\usepackage{multirow}
\begin{document}

\maketitle

% the abstract here must exactly match the abstract entered into the paper submission system
\begin{abstract}
Multilingual speaker verification remains challenging because language-dependent acoustic variability causes speaker identity to become entangled with linguistic characteristics, degrading generalization across languages.
In multilingual training, embeddings often encode language cues with speaker identity, causing speakers to form language-specific clusters.
We propose L-Proto, a language-aware episodic prototypical training strategy that constructs language-consistent episodes.
By sampling speakers from a single language per episode, L-Proto reduces language-driven variation during training and encourages embeddings to focus more directly on speaker identity.
Experiments on the TidyVoice Challenge benchmark demonstrate consistent performance improvements over conventional fine-tuning and random episodic sampling across multiple backbone architectures.
\end{abstract}

\section{Introduction}

Speaker verification (SV) aims to determine whether two speech samples belong to the same speaker.
Deep learning has significantly advanced this task by learning speaker-discriminative representations directly from speech signals~\cite{8462665, desplanques20_interspeech, jung22_interspeech}, achieving strong performance on standard benchmarks~\cite{NAGRANI2020101027, chung18b_interspeech, 10446780, lin24j_interspeech}.
However, multilingual speaker verification remains challenging.
Speaker identity is expressed through language-dependent phonetic and prosodic patterns, and training on multiple languages often causes speaker identity to become entangled with linguistic characteristics in the learned embeddings~\cite{thienpondt20_interspeech}.
As a result, utterances from the same speaker may form language-specific subclusters, and verification performance often degrades when enrollment and test utterances are in different languages or when language distributions are imbalanced~\cite{940862, 10.1007/978-3-030-71711-7_18, farhadipour2026tidyvoice, 10.1007/978-3-319-99579-3_73}.

% \vspace{+0.45cm}
To address cross-lingual variability, prior work has explored several complementary directions.
Language-adversarial approaches introduce auxiliary classifiers with gradient reversal to encourage language-invariant embeddings~\cite{8683616, 8682259, chen20n_interspeech, JMLR:v17:15-239}.
Other studies explicitly disentangle speaker and language factors through structured objectives or contrastive learning~\cite{srinivasmenon25_interspeech, olijslager2026causally}.
Domain generalization frameworks further improve robustness to language shifts using meta-learning or distributionally robust optimization~\cite{10053562, li25h_interspeech, yang22j_interspeech}, while language-aware score calibration methods have also been investigated~\cite{9746210, thienpondt20_interspeech}.
In addition, large-scale multilingual pretraining has been shown to enhance cross-lingual robustness by exposing models to diverse linguistic conditions during representation learning~\cite{lin24j_interspeech, babu22_interspeech, 9814838}.
Despite their differences, these approaches primarily operate at the representation level through global objectives applied across the full multilingual training set.
In contrast, we focus on controlling linguistic variability at the task level by shaping the language composition of episodic training.

Another line of research investigates metric-learning and episodic training strategies for SV.
Prototypical networks perform distance-based classification by forming speaker prototypes within episodic tasks~\cite{9054471}, and subsequent studies have explored improved support–query partitioning and optimization strategies for stable episodic learning~\cite{chen21f_interspeech}.
Episodic supervision enhances local discrimination among sampled speakers.
Most existing methods construct episodes by randomly sampling speakers from the training set~\cite{9054471, chen21f_interspeech}. % kumar2020designing, wang2023few
In multilingual settings, such episodes may contain speakers and utterances from multiple languages.
When languages are mixed within an episode, embeddings of the same speaker may form language-dependent sub-clusters, reflecting speaker–language entanglement.
This can distort prototype estimation and affect similarity comparisons within the episode, reducing the reliability of episodic supervision in multilingual settings.
However, how to construct episodic tasks that remain reliable under multilingual speaker–language entanglement has received relatively less attention.

These observations suggest that controlling linguistic variability at the task level can be beneficial for episodic learning in multilingual speaker verification.
To this end, we propose L-Proto, a language-aware episodic prototypical training framework that constructs language-consistent episodes.
Each episode contains speakers from a single language, reducing intra-episode linguistic variability and stabilizing prototype-based similarity learning.
Different languages are sampled across training iterations, ensuring that the model remains exposed to multilingual data throughout training.
Experimental results further suggest that L-Proto mitigates speaker–language sub-clustering, encouraging embeddings of the same speaker to remain more consistent across languages.
The main contributions of this work are summarized as follows:

\begin{itemize}
\item We show that language mixing within episodes can bias prototype estimation and destabilize similarity-based supervision in multilingual speaker verification.
\item We propose L-Proto, a language-aware episodic training strategy that constructs language-consistent episodes for multilingual speaker verification.
\item Experiments on TidyVoice demonstrate consistent gains across diverse backbones and cross-lingual scenarios, with code publicly available at \url{https://github.com/hs-oh-prml/L-Proto/}.
\end{itemize}

\begin{algorithm}[t]
\caption{Language-aware Episodic Sampling}
\label{alg:lproto}
\begin{algorithmic}[1]
\STATE Initialize buffers $\mathcal{B}[\ell][s] \leftarrow \emptyset$
\STATE Initialize ready speaker sets $\mathcal{R}[\ell] \leftarrow \emptyset$

\FOR{each incoming sample $(x, s, \ell)$}
\STATE \COMMENT{$x$: speech sample, $s$: speaker ID, $\ell$: language ID}

    \IF{$\ell$ is invalid}
        \STATE \textbf{continue}
    \ENDIF

    \STATE $\mathcal{B}[\ell][s] \leftarrow \mathcal{B}[\ell][s] \cup \{x\}$

    \IF{$|\mathcal{B}[\ell][s]| \ge K$}
        \STATE $\mathcal{R}[\ell] \leftarrow \mathcal{R}[\ell] \cup \{s\}$
    \ENDIF

    \FOR{each language $\ell'$ with $|\mathcal{R}[\ell']| \ge P$}

        \STATE Randomly sample $P$ distinct speakers $\{s_1,\dots,s_P\}$ $\subset \mathcal{R}[\ell']$
        \STATE Initialize episode $\mathcal{E} \leftarrow \emptyset$

        \FOR{each selected speaker $s_i$}

            \STATE Randomly select $K$ samples $\{x_{i,1},\dots,x_{i,K}\} \subset \mathcal{B}[\ell'][s_i]$
            \STATE $\mathcal{E} \leftarrow \mathcal{E} \cup \{(x_{i,k}, s_i)\}_{k=1}^{K}$

            \STATE $\mathcal{B}[\ell'][s_i] \leftarrow \mathcal{B}[\ell'][s_i] \setminus \{x_{i,1},\dots,x_{i,K}\}$

            \IF{$|\mathcal{B}[\ell'][s_i]| < K$}
                \STATE $\mathcal{R}[\ell'] \leftarrow \mathcal{R}[\ell'] \setminus \{s_i\}$
            \ENDIF

        \ENDFOR

        \STATE Output episode $\mathcal{E}$

    \ENDFOR

\ENDFOR
\end{algorithmic}
\end{algorithm}
% \vspace{-0.3cm}

\section{Method}

We propose L-Proto, a language-aware episodic prototypical training framework for multilingual speaker verification.
By constructing language-consistent episodes, it stabilizes prototype estimation during episodic learning.
The framework consists of language-aware episode construction, streaming episode sampling, and episodic prototypical optimization.

\subsection{Problem formulation}

Let the multilingual training set be defined as
\[
\mathcal{D} = \{(x_i, y_i, \ell_i)\}_{i=1}^N,
\]
where $x_i$ denotes an input speech sample, $y_i$ the speaker label, and $\ell_i$ the language label.
The speaker encoder $f_\theta(\cdot)$ maps an input sample to an embedding vector
\[
\mathbf{z}_i = f_\theta(x_i) \in \mathbb{R}^d.
\]

Standard training optimizes a global speaker classification objective over all samples in $\mathcal{D}$, encouraging inter-speaker separability across languages.
In contrast, episodic training introduces local discrimination among sampled speaker subsets.
L-Proto modifies this sampling procedure by enforcing language consistency within each episode.

\subsection{Language-aware episode construction}

To reduce cross-lingual variability within episodic training, we construct language-consistent episodes.
For a given language $\ell$, we sample $P$ distinct speakers and $K$ utterances per speaker to form an episode
\[
\mathcal{E}_\ell = \{(x_{(s,k)}, y_s)\},
\]
where $s \in \{1,\dots,P\}$ indexes speakers and $k \in \{1,\dots,K\}$ indexes utterances.
All samples within $\mathcal{E}_\ell$ share the same language label, so intra-episode variation is primarily determined by speaker identity rather than linguistic differences.
This reduces cross-lingual interference during prototype estimation, allowing similarity comparisons to focus more on speaker identity.
Although each episode contains a single language, episodes from different languages appear across training iterations, preserving exposure to multilingual data throughout training.
This design reduces linguistic variability within each task while still allowing the encoder to observe diverse language conditions across tasks.

\subsection{Streaming episode sampling}

To implement this episode construction strategy in practice, we employ a streaming sampling mechanism.
Each buffer $\mathcal{B}[\ell][s]$ stores utterances from speaker $s$ in language $\ell$.
When a speaker accumulates at least $K$ utterances, it becomes a ready speaker for that language.
Once at least $P$ ready speakers are available within the same language, an episode is formed by randomly selecting $P$ speakers and $K$ utterances per speaker.
Algorithm~\ref{alg:lproto} summarizes the sampling procedure.
This strategy generates language-consistent episodes while preserving stochasticity over speakers and utterances.
Episodes are formed on-the-fly during data loading without offline pre-grouping.

\subsection{Episodic prototypical objective}

Given an episode $\mathcal{E}_\ell$, embeddings are computed as $\mathbf{z} = f_\theta(x)$.
For each speaker $s$, a prototype is defined as the mean of its support embeddings:
\begin{equation}
\mathbf{p}_s =
\frac{1}{|\mathcal{S}_s|}
\sum_{\mathbf{z} \in \mathcal{S}_s} \mathbf{z},
\end{equation}
where $\mathcal{S}_s$ denotes the support set of speaker $s$.
For a query embedding $\mathbf{q}$, cosine similarity to each prototype is computed as
\begin{equation}
\mathrm{sim}(\mathbf{q}, \mathbf{p}_s)
=
\frac{\mathbf{q}^\top \mathbf{p}_s}
{|\mathbf{q}| |\mathbf{p}_s|}.
\end{equation}

The episodic loss is defined as a cross-entropy objective over similarity scores:
\begin{equation}
\mathcal{L}_{\mathrm{epi}}
=
-
\log
\frac{
\exp(\mathrm{sim}(\mathbf{q}, \mathbf{p}_{y_q}) / \tau)
}{
\sum_{s'=1}^{P}
\exp(\mathrm{sim}(\mathbf{q}, \mathbf{p}_{s'}) / \tau)
},
\end{equation}
where $y_q$ is the ground-truth speaker label of the query and $\tau$ denotes the temperature parameter used to scale cosine similarities.
This objective enforces proximity between queries and their corresponding prototypes within each episode.
Since episodes are language-consistent, similarity comparisons are less influenced by cross-lingual variability.

\begin{figure*}[t]
  \centering
  \includegraphics[width=\linewidth]{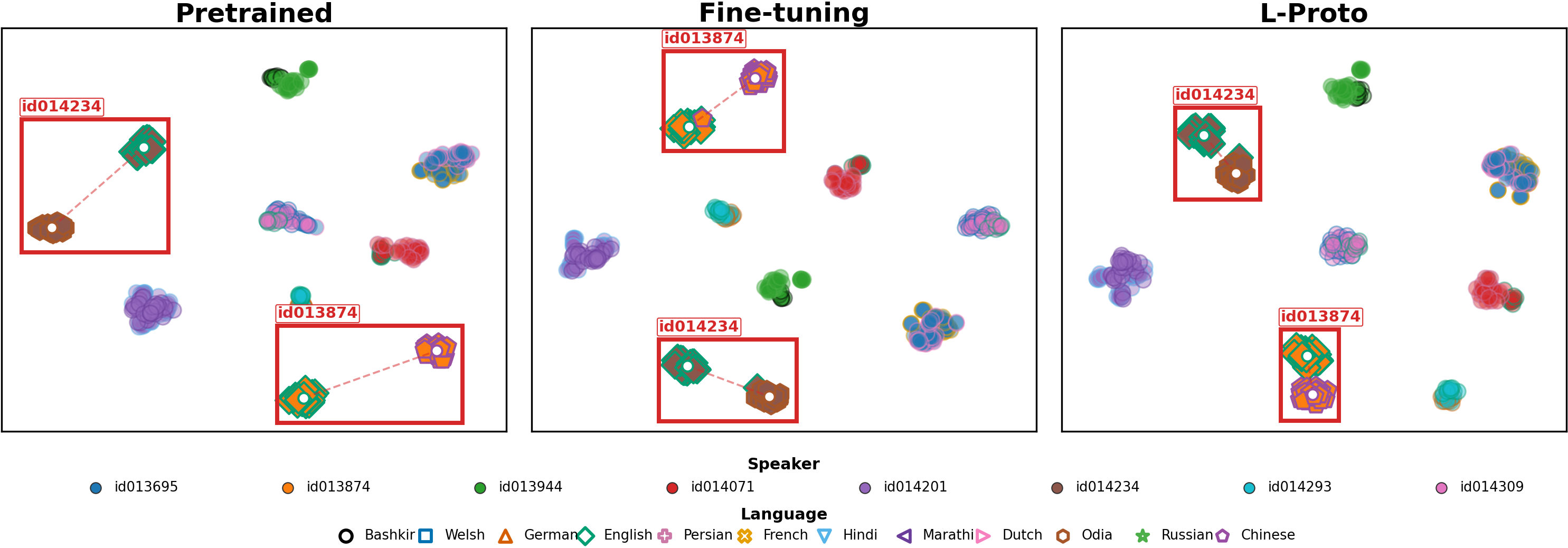}
  \caption{t-SNE visualization of embeddings for a representative speaker subset under three training settings: (a) Pretrained, (b) Fine-tuning, and (c) L-Proto.
  Points are colored by speaker identity and edge color indicates language, showing that language-dependent sub-clusters emerge under conventional fine-tuning but become more consistent across languages with L-Proto.}
  \label{fig:three}
\end{figure*} 
\vspace{-0.3cm}

\subsection{Joint training objective}

The overall training objective combines global classification and episodic supervision:
\begin{equation}
\mathcal{L}
=
\mathcal{L}_{\mathrm{cls}}
+
\lambda
\mathcal{L}_{\mathrm{epi}},
\end{equation}
where $\mathcal{L}_{\mathrm{cls}}$ denotes the speaker classification loss, and $\lambda$ controls the weight of episodic supervision.
Model parameters $\theta$ are optimized end-to-end using this joint objective.

\begin{table}[!t]
\centering
\caption{Performance on the TidyVoice Challenge development set under different language-mismatch conditions.}
\vspace{-0.2cm}
\label{tab:main_results}
\resizebox{1.00\linewidth}{!}{
\begin{tabular}{l|cccccc}
\toprule
\multirow{2}{*}{Model} &
\multicolumn{4}{c}{Target / Non-Target} &
\multicolumn{2}{c}{All} \\
\cmidrule(lr){2-5}\cmidrule(lr){6-7}
& D/D & D/S & S/D & S/S & EER & minDCF \\
\midrule
SimAM-ResNet34              & 1.56 & 4.81 & 0.95 & 2.81 & 2.88 & 0.85 \\
w/ Fine-tuning         & 1.62 & 4.94 & 0.85 & 3.03 & 2.91 & 0.81 \\
w/ L-Proto            & \textbf{0.83} & \textbf{2.24} & \textbf{0.55} & \textbf{1.69} & \textbf{1.38} & \textbf{0.63} \\
\midrule
SimAM-ResNet100             & 2.24 & 5.58 & 1.33 & 3.29 & 3.48 & 0.81 \\
w/ Fine-tuning         & 1.17 & 4.70 & 0.51 & 2.39 & 2.63 & 0.77 \\
w/ L-Proto            & \textbf{0.67} & \textbf{2.01} & \textbf{0.39} & \textbf{1.43} & \textbf{1.18} & \textbf{0.61} \\
\bottomrule
\end{tabular}
}
% \vspace{-0.3cm}
\end{table}

\begin{table}[!t]
\centering
\caption{Comparison across public speaker embedding backbones: pretrained, fine-tuned, and L-Proto models.} 
\vspace{-0.2cm}
\label{tab:exp_baselines}
\resizebox{1.00\linewidth}{!}{
\begin{tabular}{l|cc|cc|cc}
\toprule

\multirow{2}{*}{Model} &
\multicolumn{2}{c|}{Pretrain} &
\multicolumn{2}{c|}{Fine-tuning} &
\multicolumn{2}{c}{L-Proto} \\
&
{EER} & {minDCF} &
{EER} & {minDCF} &
{EER} & {minDCF} \\
\midrule
ResNet34                    & 3.23 & 0.88 & 3.09 & 0.82 & \textbf{2.58} & \textbf{0.78} \\
ResNet152                   & 2.94 & 0.88 & 3.27 & 0.85 & \textbf{2.49} & \textbf{0.74} \\
ResNet221                   & 3.32 & 0.92 & 3.28 & 0.86 & \textbf{2.61} & \textbf{0.76} \\
ResNet293                   & 3.04 & 0.86 & 3.28 & 0.85 & \textbf{2.44} & \textbf{0.75} \\
ECAPA512                    & 4.15 & 0.96 & 4.52 & \textbf{0.92} & \textbf{3.73} & 0.95 \\
ECAPA1024                   & 4.70 & 1.00 & 4.95 & 0.96 & \textbf{3.69} & \textbf{0.92} \\
CAM++                       & 4.06 & 0.93 & 3.90 & 0.92 & \textbf{2.50} & \textbf{0.76} \\
\bottomrule
\end{tabular}
}
% \vspace{-0.3cm}
\end{table}

\section{Experiments}
\subsection{Experimental setup}

We evaluate SimAM-ResNet34 and SimAM-ResNet100~\cite{9746294}, ResNet-based speaker embedding encoders with simple attention modules, using the \textit{wespeaker} toolkit~\cite{wang2024advancing}.
The models are initialized from checkpoints pretrained on VoxBlink2~\cite{lin24j_interspeech}, fine-tuned on TidyVoiceX~\cite{farhadipour2026tidyvoice}, and evaluated on the official TidyVoice Challenge development set.
TidyVoiceX contains over 4,474 speakers, around 40 languages, and approximately 321k utterances from Mozilla Common Voice.
The challenge evaluates similarity scores on same- and different-speaker trials under same- and cross-lingual conditions.
We report EER and minDCF on the development set because official evaluation labels are not publicly available.
Fine-tuning is performed for 6 epochs following the standard configuration used in the TidyVoice Challenge baseline setup.
The learning rate is decayed from $5\times10^{-5}$ to $1\times10^{-5}$. 
Input features are 80-dim log Mel-filterbanks extracted from 16~kHz audio with 600-frame random cropping.
For episodic training, each episode contains $P$ speakers with two utterances per speaker.
We set $P=24$ and $\lambda=100$ for ResNet34, and $P=8$ and $\lambda=200$ for ResNet100, with a temperature of 0.07.
Hyperparameters were selected on the development set.
Experiments are conducted on two NVIDIA RTX A6000 GPUs.
All models are trained under the same data split and evaluation protocol for fair comparison.

\subsection{Overall performance}

Table~\ref{tab:main_results} summarizes performance under four trial language conditions.
S and D denote same- and different-language pairs, respectively.
Across both backbone architectures, L-Proto improves EER and minDCF over pretrained and fine-tuned baselines.
The gains are most pronounced when trials involve cross-lingual target pairs (D/D and D/S), where cross-lingual variability directly affects similarity estimation.
By constructing language-consistent episodes, L-Proto reduces language-induced variation in prototype-based discrimination while preserving multilingual coverage during training.
L-Proto also improves matched-language conditions (S/D and S/S), indicating better speaker separability without degrading same-language performance.

\begin{table}[t]
\centering
\caption{Centroid cosine similarity analysis.
$\mathrm{Sim}{\text{intra}}$ measures cross-lingual centroid similarity within the same speaker, while $\mathrm{Sim}{\text{inter}}$ measures same-language centroid similarity across different speakers. $\Delta$ denotes their separation margin.} 

% \vspace{-0.3cm}
\label{tab:centroid_sim}
\resizebox{0.75\linewidth}{!}{
\begin{tabular}{l|ccc}
\toprule 
Method & $\mathrm{Sim}_{\text{intra}}\uparrow$ & $\mathrm{Sim}_{\text{inter}}\downarrow$ & $\Delta\uparrow$ \\
\midrule
Pretrained   & 0.8027 & 0.2477 & 0.5550 \\
Fine-tuning  & 0.7892 & 0.1048 & 0.6844 \\
L-Proto      & \textbf{0.8536} & \textbf{0.0479} & \textbf{0.8057} \\
\bottomrule
\end{tabular} 
}
\vspace{-0.4cm}
\end{table}

\subsection{Generalization across backbone architectures}

Table~\ref{tab:exp_baselines} evaluates L-Proto on a diverse set of publicly available pretrained speaker embedding models\footnote{\url{https://github.com/areffarhadi/wespeaker/blob/master/docs/pretrained.md}}.
They include VoxCeleb-pretrained ResNet variants~\cite{He_2016_CVPR}, CAM++~\cite{wang23ha_interspeech}, and ECAPA-TDNN variants~\cite{desplanques20_interspeech}.
For each model, we compare the original pretrained checkpoint, conventional fine-tuning on TidyVoiceX, and fine-tuning with L-Proto.
Across all evaluated backbones, L-Proto yields the lowest EER.
This suggests that language-aware episode construction is not tied to a particular architecture.
For minDCF, L-Proto also yields lower values in most cases, indicating improved calibration of similarity scores under language mismatch.
The magnitude of improvement varies across architectures, likely reflecting differences in model capacity and pretraining conditions.

\subsection{Speaker–language disentanglement}

Figure~\ref{fig:three} visualizes embedding distributions under different training strategies. 
Points are colored by speaker identity, while edge colors indicate language. 
In pretrained and fine-tuned models, utterances from the same speaker tend to form language-specific sub-clusters. 
In contrast, L-Proto produces more coherent clusters across languages, reflecting improved cross-lingual consistency. 
The red boxes illustrate that L-Proto merges language-specific clusters of the same speaker more effectively.
To quantify this observation, we analyze cosine similarities between speaker--language centroids as summarized in Table~\ref{tab:centroid_sim}. 
We compute $\mathrm{Sim}_{\text{intra}}$ between language-specific centroids of the same speaker and $\mathrm{Sim}_{\text{inter}}$ between different speakers within the same language. 
L-Proto increases intra-speaker cross-lingual similarity while reducing inter-speaker similarity, resulting in a larger separation margin.

\begin{table}[t]
  \centering
  \caption{Effect of episodic sampling and prototype supervision.}
  \vspace{-0.2cm}
  \label{tab:ablation}
  \resizebox{0.71\linewidth}{!}{
  \renewcommand{\arraystretch}{1.1}
  \begin{tabular}{cc|cc}
    \toprule 
    Episode & Prototype & EER & minDCF \\
    \midrule
    \texttimes & \texttimes & 3.48  & 0.81 \\
    \checkmark & \texttimes & 2.40  & 0.74 \\
    \texttimes & \checkmark & 2.63  & 0.77 \\
    \checkmark & \checkmark & \textbf{1.18}  & \textbf{0.61} \\
    \bottomrule
  \end{tabular}
  }
  % \vspace{-0.3cm}
\end{table}

\begin{table}[!t]
\centering
\caption{Sensitivity to the episodic loss weight $\lambda$.}
\vspace{-0.2cm}
\label{tab:protoloss}
\resizebox{0.75\linewidth}{!}{
\begin{tabular}{l|c|cc}
\toprule
Method & $\lambda$ & EER & minDCF \\
\midrule
\multirow{4}{*}{SimAM-ResNet34} & 0   & 2.70 & 0.76 \\
              & 50  & 1.45 & 0.64 \\
              & 100 & \textbf{1.38} & \textbf{0.63} \\
              & 200 & 1.46 & 0.64 \\
\midrule
\multirow{4}{*}{SimAM-ResNet100} & 0   & 2.17 & 0.71 \\
               & 100 & 1.25 & 0.62 \\
               & 200 & \textbf{1.18} & \textbf{0.61} \\
               & 300 & 1.21 & 0.62 \\
\bottomrule
\end{tabular} 
}
% \vspace{-0.3cm}
\end{table}

\begin{table}[!t]
\centering
\caption{Effect of episode construction strategy.} 
\vspace{-0.2cm}
\label{tab:lproto}
\resizebox{0.70\linewidth}{!}{
\begin{tabular}{l|cc}
\toprule
Method & EER & minDCF \\
\midrule
Random sampling & 1.54 & 0.66 \\
2 languages     & 1.35 & 0.65 \\
4 languages     & 1.69 & 0.69 \\
1 language      & \textbf{1.18} & \textbf{0.61} \\
\bottomrule
\end{tabular} 
}
\vspace{-0.3cm}
\end{table}

\subsection{Ablation study}

In Table~\ref{tab:ablation}, the setting without episodic sampling and prototype supervision corresponds to conventional fine-tuning with only the global classification objective.
Table~\ref{tab:ablation} evaluates episodic sampling and prototype-based supervision. 
Each component improves the baseline, and their combination gives the largest gain.
Table~\ref{tab:protoloss} examines the effect of the episodic loss weight $\lambda$. Using the episodic loss improves over $\lambda=0$ and remains stable across values.
Table~\ref{tab:lproto} analyzes the impact of episode language composition. 
Multi-language episodes do not improve consistently and can degrade as language diversity increases, suggesting that mixed-language episodes introduce nuisance variation into prototype estimation.
Restricting each episode to a single language yields the best results, showing that L-Proto benefits from both episodic supervision and language-consistent episode construction.
This trend is consistent with our motivation that reducing intra-episode linguistic variability stabilizes prototype-based supervision.

\subsection{Language-wise improvement over pretrained models}

\begin{figure}[!t]
  \centering
  \includegraphics[width=1.0\linewidth] {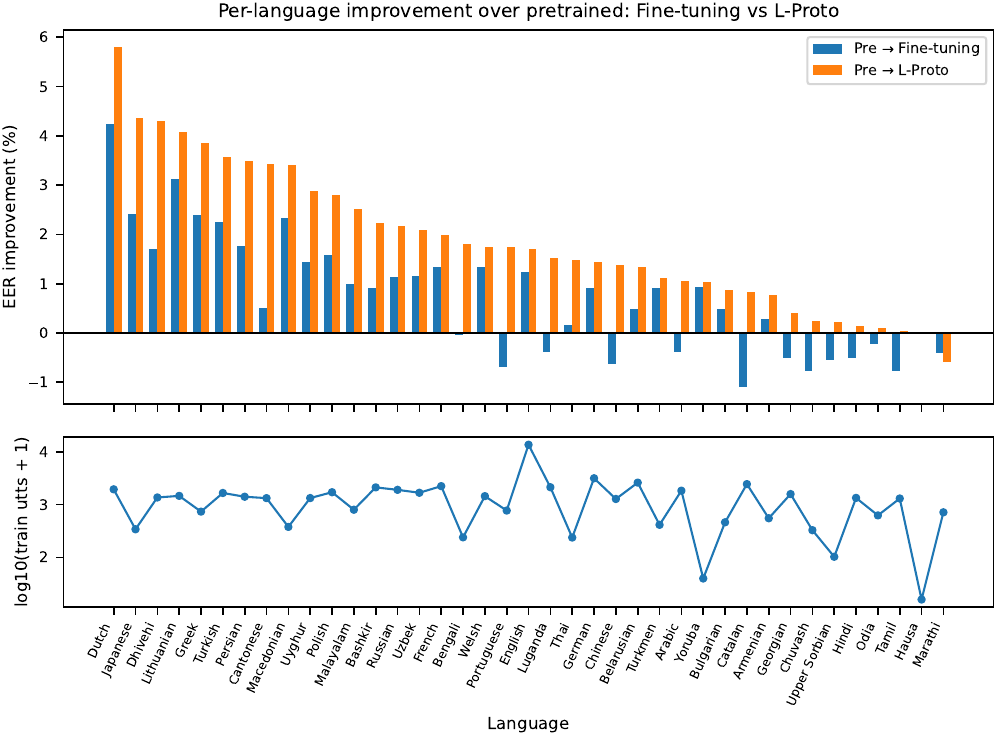} 
  \vspace{-0.2cm}
  \caption{Language-wise improvement over the pretrained model. Top: EER reduction. Bottom: training data size.}
  \vspace{-0.2cm}
  \label{fig:data_analysis}
\end{figure}

Figure~\ref{fig:data_analysis} compares the improvement in EER over the pretrained model for each language in the development set. 
The upper panel shows the EER reduction achieved by fine-tuning and by the proposed L-Proto training, while the lower panel indicates the amount of available training data for each language.
Both fine-tuning and L-Proto substantially improve over the pretrained model across languages. 
L-Proto generally yields larger gains than fine-tuning, suggesting better adaptation to multilingual conditions.
The language-wise results show that L-Proto improves most languages, although the magnitude of improvement varies, indicating that its benefit is not uniform across language conditions.

\section{Conclusion}

This paper introduced L-Proto, a language-aware episodic prototypical training strategy for multilingual speaker verification.
By constructing single-language episodes, L-Proto reduces intra-episode linguistic variation and stabilizes prototype-based similarity learning.
Experiments on the TidyVoice Challenge benchmark show consistent improvements over fine-tuning and random episodic sampling across multiple backbones.
The method requires language labels and sufficient speaker diversity within each language, and introduces additional sampling overhead.
Future work will explore adaptive episode construction for balancing linguistic diversity and speaker discrimination.

\section{Use of Generative AI Disclosure}

Generative AI tools (OpenAI GPT-5.2) were used for writing assistance, including language editing and minor revisions of the manuscript. All technical content and conclusions were developed by the authors.

\section{Acknowledgments}
This work was partly supported by Institute of Information \& Communications Technology Planning \& Evaluation (IITP) grants funded by the Korea government (MSIT) (No. RS-2019-II190079, Artificial Intelligence Graduate School Program (Korea University); No. RS-2024-00336673, AI Technology for Interactive Communication of Language Impaired Individuals; and No. IITP-2026-RS-2025-02304828, Artificial Intelligence Star Fellowship Support Program to nurture the best talents). 

\bibliographystyle{IEEEtran}
\bibliography{refs}

\end{document}